\documentclass[pra,showpacs,twocolumn]{revtex4-1}
\usepackage{bm,graphicx,amsmath, color}

\newcommand{\beq}{\begin{equation}}
\newcommand{\enq}{\end{equation}}

\begin{document}
\title{$p$ orbitals in 3D lattices; fermions, bosons and (exotic) models of
magnetism}  
\author{Fernanda Pinheiro} 
\email{fep@fysik.su.se}  
\affiliation{Department of Physics, Stockholm University, Se-106 91
  Stockholm, Sweden} 
\affiliation{NORDITA,
KTH Royal Institute of Technology and Stockholm University, Se-106 91
Stockholm, Sweden}  
\date{\today}

\begin{abstract}
We demonstrate how different types of $SU(3)$ Heisenberg models can
be implemented with the use of the $p$ orbitals of
three dimensional optical lattices. By considering a Mott insulator
with unit filling, the dynamics is well described by
an effective model derived from the perturbative treatment of the
tunneling elements 
relative to the onsite interaction terms. 
This yields systems with degrees of freedom that are generators of
the $SU(3)$ group, which 
extends the Heisenberg models frequently used to analyze
quantum magnetism.  
Due to the 
different character of interactions  
in the bosonic and fermionic cases, the choice of
atom determines what 
type of
anisotropies will appear in the couplings of the 
corresponding effective 
Hamiltonians. Experimental schemes for detection and manipulation of
these systems are presented, and properties of the
ground states of selected examples are discussed. 

\end{abstract}
\pacs{03.75.Lm, 67.85.Hj, 05.30.Rt}
\maketitle

\section{Introduction}
The amazing degree of control and manipulation in experiments with
systems of cold atoms has recently awaken a renewal of interest in the
physics of interacting spins beyond spin $1/2$~\cite{toth2010three,
  Bauer:2012aa, Li:1998aa}. This follows
the success with implementation of (spin-$1/2$) models of magnetism in
the lab~\cite{simon2011quantum, struck2013engineering,
  krauser2014giant, ebling2014relaxation}, and opens a great window
for application of 
quantum simulators.  A question of great relevance that therefore
arises is: what are the experimentally feasible many-body systems that
can be used to mimic the physics of particular models of interacting
spins? (And how?)

Along these lines, realizations of $SU(N)$ Heisenberg models have been
proposed with fermionic atoms in optical lattices~\cite{toth2010three,
  Bauer:2012aa, Li:1998aa}. In these
systems, the pseudospin degree of freedom is encoded in the nuclear spin
states, whose number $N$ determines the $SU(N)$ symmetry of the
corresponding model~\cite{honerkamp2004ultracold}. Furthermore, different
setups have been suggested 
with trapped ions for the implementation of the spin $1$ $XY$
Hamiltonian~\cite{senko2014experimental}, and of $SU(3)$ chains with
long-range 
interaction~\cite{grass2013quantum}. Realizations of $SU(3)$
Heisenberg models with bosonic 
Mott insulators with spin-orbit coupling  have also been
proposed~\cite{grass2014spiral}. 

In this paper, we present an alternative controllable system for
realization of 
different types of $SU(3)$ Heisenberg models. We show
that the effective dynamics of the Mott phase
with one atom per site (Mott$_1$) in the first excited bands 
of $3D$ optical lattices is equivalent to that of a system of nearest
neighbours interactions, where the degrees of freedom are the
generators of the $SU(3)$ group. Rather than internal electronic
atomic states, the pseudospin is encoded here in the vibrational states
which correspond to the (onsite) $p$ orbitals of the optical potential. 
In $3D$, the first excited bands, or $p$ band, can be made 3-fold
degenerate. This is the case, for example, in the cubic
lattice, where 
the three orbital states, generally called $p_x$, $p_y$ and $p_z$
orbitals, are anisotropic both in magnitude and
parity~\cite{isacsson2005multiflavor, lewenstein2011optical}. As a
consequence, the dynamics of many-body systems in
the $p$ band can be considerably different than multispecies systems
in the 
ground or $s$ band. It is characterized by anisotropic tunneling
amplitudes, and
in addition to density-density 
interactions, the bosonic case features processes that transfer
population between the different orbital states. 
These additional properties have already been shown to have important
effects for the bosonic Mott phase with one particle per site in $2D$
optical lattices, for its effective dynamics corresponds to that of
the fully anisotropic spin-$1/2$ Heisenberg model in an external
field~\cite{pinheiro2013xyz}.  

Our focus here is the physics of the Mott$_1$ phase in the 
$p$ band of $3D$ optical lattices, which is 
$3$-fold degenerate (or quasi-degenerate).
We study different systems in terms of their corresponding effective
Hamiltonians, obtained from the perturbative treatment of 
tunneling elements up to second order. We will extend the method used
in Ref.~\cite{pinheiro2013xyz} to account for 
the $3$-orbital case and we discuss properties of both the fermionic
and the bosonic cases. As will be shown, the statistics of the atoms
play an important role in determining the anisotropies in the
couplings of the effective Hamiltonians. We will also characterize the
ground states in of the some specific cases. 
For antiferromagnetic 
$SU(3)$ Heisenberg models, very rich physics is expected from
competition between quantum and 
classical fluctuations~\cite{toth2010three, Bauer:2012aa}, which lead
to the formation of ground states 
with 2 and 3-sublattice ordering both in $2D$ and
$3D$~\cite{papanicolaou1988unusual, toth2010three, Bauer:2012aa}. In
the ferromagnetic case, we look for ground states which feature
spiral spin textures, as recently reported in Ref.~\cite{grass2014spiral}. 

The paper is organized as follows: In Sec.~\ref{sec:dynamics_p_band}
we discuss the Hamiltonians of both the many-body bosonic and
fermionic systems in the $p$-band. We then derive the effective
Hamiltonians for the Mott$_1$ phase in Sec.~\ref{sec:eff_ham} and
discuss properties of the ground states in some special
cases. Sec.~\ref{sec:exp_prob} is devoted to experimental probing
of these systems, and in Sec.~\ref{sec:imperfec} we discuss the effects
of imperfections in experimental realization. Final remarks and
further extensions of the techniques considered here are discussed
in the Conclusions.  


\section{Dynamics in the $p$ band}\label{sec:dynamics_p_band}
We start with the general expression of the many-body Hamiltonian in
terms of the 
field operators,
\begin{equation}\label{many-body_Ham}
\begin{array}{rcl}
H &=& \displaystyle{\int d\vec r\bigg\{\hat \Psi^\dagger(\vec r) \left[
     -\frac{\hbar^2}{2m}\nabla^2 + V(\vec r)\right]\hat \Psi(\vec r) }\\ \\
& & \displaystyle{+ \frac{U_0}{2}\hat \Psi^\dagger(\vec r)\hat \Psi^\dagger(\vec
r)\hat \Psi(\vec r)\hat \Psi(\vec r) \bigg\} },
\end{array}
\end{equation}
where $m$ is the mass of the atoms, $V(\vec r)$ accounts for the
effects of external potentials in the system and $U_0 =
4\pi\hbar^2a/m$, with $a$ the s-wave scattering length, 
measures the strength of (short-range) contact interactions.
The field operators $\hat \Psi(\vec r)$ and $\hat  
\Psi^\dagger(\vec r)$ annihilate and create a particle at the position
$\vec r$, and obey commutation relations $[\hat
  \Psi(\vec r), \Psi^\dagger(\vec r^{'}) ] = \delta(r -
  r^{'})$ and anti-commutation relations and  $\{\hat
  \Psi(\vec r), \Psi^\dagger(\vec r^{'}) \} = \delta(r -
  r^{'})$ for bosons and fermions, respectively. For the periodic
  lattice considered here the dynamics is separable in the different
  directions, i.e.,
\begin{equation}\label{per_pot}
V(\vec r) = V_x\sin^2(k_xx) + V_y\sin^2(k_y y) + V_z\sin^2(k_zz),
\end{equation}
where $V_\sigma$ are the amplitudes of the lasers in the direction
$\sigma = \{x, y, x\}$ and $k_\sigma = 2\pi/\lambda_\sigma$ are the
lasers wave vectors, with $\lambda_\sigma$ the lasers
wavelengths. All energies are scaled with the recoil energy, 
$E_R = \hbar^2k_{\sigma^{'}}^2/2m$, for the $\sigma^{'}$ that defines
the smallest $k_\sigma$. 

Following the usual prescription, we study the physics in the first
excited bands of this lattice by expanding the field operators in 
(\ref{many-body_Ham}) in terms of the site-localized $p$ orbitals (or
$p$ bands)~\cite{isacsson2005multiflavor, pinheiro2012confined},
\begin{equation}\label{orbital_states}
\begin{array}{rcl}
\displaystyle{\hat \Psi(\vec r)} &=& \displaystyle{\sum_{\bf
    i}\sum_{\alpha} \psi_{\alpha, \bm i}(\vec r)\hat 
a_{\alpha, \bf i}},
\end{array}
\end{equation}
where $\alpha = \{x, y, z\}$ and $\hat a_{\alpha, \bm i}$
annihilates an atom in the $p_{\alpha}$-orbital state at the
site ${\bf i} = (i_x, i_y, i_z)$. 


\subsection{The bosonic case}
After expanding the field operators according to
(\ref{orbital_states}), and truncating the kinetic terms to its
leading contribution, i.e. the tight-binding approximation, 
the Hamiltonian describing 
bosonic atoms in the $p$-band of a $3D$ optical lattice is given by 
\begin{equation}\label{eq1}
\hat H_B = \hat H_0 + \hat H_{nn} + \hat H_{nn^{'}} + \hat H_{OD}.
\end{equation}
The first term is the free Hamiltonian
\begin{equation}\label{H_free}
\hat H_0 = \displaystyle{-\sum_{\sigma, \alpha}\sum_{\langle \bf{i},
    \bf{j}\rangle_{\sigma}}t^\alpha_{\bf ij}(\hat a^{\dagger}_{\alpha,
  \bf i}\hat
a_{\alpha, \bf j} + \hat a^{\dagger}_{\alpha, \bf j}\hat a_{\alpha, \bf i}),}
\end{equation}
that describes the nearest neighbour tunneling of atoms in the
$p_\alpha$-orbital state, 
$\alpha = \{x, y, z\}$, in the direction
$\sigma = \{x, y, z\}$.
The second and the third terms describe different types of density-density
interactions:
\begin{equation}\label{H_nn}
\hat H_{nn} = \displaystyle{\sum_{\alpha}\sum_{\bf
    i}\frac{U_{\alpha\alpha}}{2}\hat n_{\alpha, 
  \bf i}(\hat
n_{\alpha, \bf i} - 1)},
\end{equation}
beween atoms in the same orbital state,
with $\hat n_{\alpha, \bf i} = \hat a^\dagger_{\alpha, i}\hat a_{\alpha, i}$; and 
\begin{equation}\label{H_mn}
\hat H_{nn^{'}} = \displaystyle{\sum_{\alpha\neq\beta}\sum_{\bf
    i}2U_{\alpha\beta}\hat 
n_{\alpha, \bf i}\hat n_{\beta, \bf i}},
\end{equation}
$\beta = \{x, y, z\}$, between atoms in different orbital states. The last term
\begin{equation}\label{H_FD}
\hat H_{OD} =\displaystyle{\sum_{\alpha\neq\beta}\sum_{\bf
    i}\frac{U_{\alpha\beta}}{2}(\hat 
a^{\dagger}_{\alpha, \bf {i}}\hat a^{\dagger}_{\alpha, \bf {i}}\hat 
a_{\beta,\bf { i}}\hat
a_{\beta, \bf {i}} + \hat
a^{\dagger}_{\beta, \bf{i}}\hat a^{\dagger}_{\beta, \bf{i}}\hat a_{\alpha, \bf{i}}\hat
a_{\alpha, i})}
\end{equation}
describes interactions that transfer population  
within different types of orbital states. This is a key ingredient for
characterizing collective properties of bosons in the $p$ band. It
reduces the $U(1)\times U(1)\times U(1)$ global symmetry of the model
to $U(1)\times Z_2\times Z_2$, and as a consequence, total population
of each of the orbital states is conserved only modulo 2. It has also
fundamental implications, for it prevents the Hohenberg-Mermin-Wagner
theorem~\cite{Mermin:1966aa, Hohenberg:1967aa} from 
prohibiting long-range phase coherence in
low dimensional systems of bosons in the $p$ band.

In terms of the orbital states, the expression for the tunneling
coefficients in the direction $\sigma$ is given by
\begin{equation}\label{tunneling_coef}
t^{\alpha}_{\bf ij} = \displaystyle{-\int d\vec r \,\psi^{*}_{\alpha \bf i}(\vec
r)\left[-\nabla^2 + V(\vec r) \right]\psi_{\alpha \bf{j +
    1}_{\sigma}}(\vec r) },
\end{equation}
and due to the differences in the curvature of the excited bands in the
directions perpendicular and parallel to the label of the orbital wave
functions, 
$t_{\perp}t_{\parallel} < 0$, where $t_{\perp}$  and $t_{\parallel}$
refer, respectively, to the perpendicular and parallel
tunnelings. Notice furthermore that the label of the orbital state
is a reference to the direction which introduces the node in the orbital
wavefunction. In the same way, the expression of the interaction coefficients is 
given by
\begin{equation}\label{interaction_coef}
U_{\alpha\beta} = \displaystyle{U_0\int d\vec r\,|\psi_{\alpha \bf j}(\vec r)
  |^2 |\psi_{\beta \bf j}(\vec r) |^2 }.
\end{equation}
As final remark we recall that  in the bosonic case $[\hat a_{\alpha, \bf i},
\hat a_{\beta, \bf j}] = \delta_{\alpha\beta}\delta_{{\bf i}, {\bf j}}$. 

\subsection{The fermionic case}
Anti-commutation relations prevent the occupation of the same orbital
state by more than one identical fermion. Therefore, (spinless)
fermionic atoms in the $p$ band behave according to 
\begin{equation}\label{fermions_mb}
\hat H_F = \hat H_0 + \hat H_{nn^{'}},
\end{equation}
with $\hat H_0$ and $\hat H_{nn^{'}}$ defined in
Eqs.~(\ref{H_free}) and~(\ref{H_mn}), respectively. Here, however,
$\{\hat a_{\alpha, \bf i}, \hat a_{\beta, \bf j}\} =
\delta_{\alpha\beta}\delta_{{\bf i}, {\bf j}}$. The expressions for the
tunneling elements and the various coupling contants are the same as
in the bosonic case, defined in Eqs.~(\ref{tunneling_coef})
and~(\ref{interaction_coef}).

\section{Effective Hamiltonian}\label{sec:eff_ham}
We now turn to the physics deep in the Mott$_1$ phase. 
In this regime $|t^{\alpha\alpha}| \ll U_{\alpha\beta}$
and   
the dynamics can be effectively described by a Hamiltonian where 
the tunneling processes are treated/included
perturbatively~\cite{assa1994interacting}. 
This effective Hamiltonian is
most easily derived by 
introducing the $\hat P$ and $\hat Q$ projectors, $\hat P + \hat Q =
1$, $\hat P^2 = \hat P$ and $\hat Q^2 = \hat Q$, that project, 
respectively, into Hilbert space of states with singly occupied sites,
$\mathcal{H_P}$, 
and the states that have at least one site with double occupation,
$\mathcal{H_Q}$. The 
eigenvalue problem $\hat H\Psi = E\Psi$ can then be  
written as 
\begin{equation}\label{eq7}
\hat H (\hat P + \hat Q)\Psi \rightarrow (\hat H_K + \hat H_U)(\hat P
+ \hat Q)\Psi = E\Psi,
\end{equation}
where $\hat H_K = \hat H_0$ is the kinetic part of  Eq.~(\ref{eq1}), and
$\hat H_U$ is the interaction Hamiltonian. Now
acting from the left with the projectors on Eq.~(\ref{eq7}), 
\begin{equation}
\left(\hat Q\hat H_K \hat P + \hat Q \hat H_K \hat Q + \hat Q \hat
  H_U\hat P + \hat Q \hat H_U\hat Q \right)\Psi =
E\hat Q\Psi,
\end{equation}
\begin{equation}
\left(\hat P\hat H_K\hat P + \hat P \hat H_K\hat Q + \hat P \hat
  H_U\hat Q + \hat Q\hat H_U\hat P\right)\Psi=  E\hat P\Psi.
\end{equation}
$\hat Q\hat H_K\hat Q$ and $\hat
P\hat H_U\hat P$ are identically zero for computing overlaps between
elements 
projected in disjoint subspaces of the Hilbert
space. $\hat P\hat H_U\hat P$ is also zero
because it computes interactions in singly occupied
sites. We then obtain
\begin{equation}
\hat Q\Psi = -\frac{1}{\hat Q\hat H\hat Q - E}\hat Q\hat H_K\hat P\Psi,
\end{equation}
which leads to
\begin{equation}\label{H_eff}
\hat H_{Mott_1} = -\hat P\hat H_K\hat Q \frac{1}{\hat Q \hat H \hat Q
  - E}\hat Q\hat H_K \hat P.
\end{equation}
So far this expression is free from approximations and will serve as the
starting point in the derivation of the effective Hamiltonian describing the
Mott$_1$ phase of the systems considered here. This language is
particularly useful for highlighting the role of the tunneling elements, namely
of connecting the intermediate and final states of the perturbative
process in the $\mathcal{H}_Q$ and $\mathcal{H}_P$ subspaces,
respectively.

Now the assumption of a Mott phase justify expanding the 
operator $1/(\hat Q\hat H
\hat Q - E)$ to second order in $t/U$, and due to the tight-binding
regime it is enough to consider the 2-site problem. We
then define the basis 
spanning the subspace of states with unit filling as
\begin{equation}
\begin{array}{rcl}
\mathcal{H}_P &=& \big\{\,\vert x, x\rangle, \vert x, y\rangle, \vert x,
z\rangle,  \vert x, y\rangle, \vert y, y\rangle, \vert y, z\rangle,
\vert z, x\rangle, \\[0.5em]
 & & \,\,\,\,\vert z, y\rangle, \vert z,
z\rangle\big\},
\end{array}
\end{equation}
where $\vert \alpha, \beta\rangle = \hat a^{\dagger}_{\alpha, \bf
  i}\hat a^{\dagger}_{\beta, \bf j}\vert 0\rangle$,
corresponds to the state  
with a $p_{\alpha}$ orbital in the site $\bf i$ and a
$p_{\beta}$ orbital in the neighbouring site $\bf j$, $\alpha, \beta =
\{x, y, z\}$.

In the same way, the relevant states in the basis of the
subspace of doubly occupied sites follow
\begin{equation}\label{H_Q}
\mathcal{H_Q} = \{\vert 0, 2x\rangle, \vert 0, 2y\rangle, \vert 0,
2z\rangle, \vert 0, xy\rangle, \vert 0, xz\rangle, \vert 0, yz\rangle\},
\end{equation}
with $\vert 0, 2\alpha\rangle = 2^{-1/2}\hat a^\dagger_{\alpha, \bf
  j}\hat a^\dagger_{\alpha, \bf j} \vert 0\rangle$ and $\vert 0, \alpha\beta\rangle
= \hat a^\dagger_{\alpha, \bf j} \hat a^\dagger_{\beta, \bf j}\vert 0
\rangle$. Notice that due to the restrictions imposed by the
exclusion principle, the relevant part of the basis in the
$\mathcal{H}_Q$ subspace 
of the fermionic problem is only composed of the  $\vert 0,
\alpha\beta\rangle 
= \hat a^\dagger_{\alpha, \bf j} \hat a^\dagger_{\beta, \bf j}\vert 0
\rangle$ states with $\alpha\neq\beta$. This leads to important differences
when deriving the effective Hamiltonians of the bosonic and fermionic
cases and will be discussed in greater detail in the following sections.

\subsection{The effective Hamiltonian for bosons}
Due to the nonvanishing contributions of the orbital changing
processes in the states of the type $\vert 0, 2\alpha\rangle =
2^{-1/2}\hat a^\dagger_{\alpha, \bf j}\hat a^\dagger_{\alpha, \bf j} \vert
0\rangle$ in the bosonic case, the projection of the
Hamiltonian in the $\mathcal{H}_Q$ subspace is non-diagonal in the
basis of intermediate states of the perturbation theory. 
As a consequence, we obtain $(\hat H_Q - E)^{-1}$, with $\hat H_Q =
\hat Q\hat H\hat Q$ by first projecting the Hamiltonian in the
$\mathcal{H}_Q$ subspace, and then taking the contributions for the
corresponding processes from the inverse of the resulting matrix. In
addition, since $E \sim t^2/U$ we take $(\hat H_Q - E)^{-1} \approx
\hat H_Q^{-1}$.

With the basis of $\mathcal{H}_Q$ ordered according to~(\ref{H_Q}),
the projected Hamiltonian
$\hat H_Q$ can be written in block diagonal form of 3x3 matrices as
\begin{equation}
H_Q = \left(\begin{array}{cc}
H_{Q_1} & 0\\
0 & H_{Q_2}\end{array}\right),
\end{equation}
where the first block captures the action of $\hat
H_U = \hat H_{nn} + \hat H_{nn^{'}} + \hat H_{OD}$ (recall the
definitions in Eqs.~(\ref{H_nn}),~(\ref{H_mn}) and~(\ref{H_FD})) in
the states of the type $\vert 0, 2\alpha\rangle$, while the second
block accounts for the effects of $\hat H_U$ in the $\vert 0,
\alpha\beta\rangle$ states. The
explicit expressions follow,
\begin{equation}
H_{Q_1} = \left( \begin{array}{ccc}
U_{xx} & U_{xy} & U_{xz}\\
U_{xy} & U_{yy} & U_{yz}\\
U_{xz} & U_{yz} & U_{zz}
\end{array}\right)
\end{equation}
and
\begin{equation}\label{K2}
H_{Q_2} = \left( \begin{array}{ccc}
2U_{xy} & 0 & 0\\
0& 2U_{xz} & 0\\
0 & 0 & 2U_{yz}
\end{array}\right),
\end{equation}
from which $\hat K = \hat H_Q^{-1}$ is easily computed. In
the first block, the elements of $\hat K^{(1)} = \hat H^{-1}_{Q_1}$
can be written as 
\begin{equation}
\begin{array}{lcr}
K^{(1)}_{\alpha\alpha} &=& \frac{1}{2\Lambda}\sum_{\beta\gamma}\left(
  \epsilon^{\alpha\beta\gamma}\right)^2\left( U_{\beta\beta}
  U_{\gamma\gamma} - U^2_{\beta\gamma}\right), \\\\
K^{(1)}_{\alpha\beta} &=&
\frac{1}{\Lambda}\sum_{\gamma}\left(\epsilon^{\alpha\beta\gamma}
\right)^ 2 \left(U_{\alpha\beta}U_{\beta\gamma} -
  U_{\alpha\beta}U_{\gamma\gamma}\right)
\end{array}
\end{equation}
where $\epsilon^{\alpha\beta\gamma}$ is the Levi-Civita symbol and
$\{\alpha, \beta, \gamma\} = (1, 2, 3)$ whenever $\{\alpha, \beta,
\gamma\} = (x, y, z)$, and 
\begin{equation}
\begin{array}{lcr}
\Lambda &=& \left(U_{xx}U_{yy}U_{zz} - U^2_{xz}U_{yy} - U^2_{yz}U_{xx} -
U^2_{xy}U_{zz}\right.\\[0.8em]
& & \left.+\, 2U_{xy}U_{xz}U_{yz}\right)^{-1}.
\end{array}
\end{equation}
For simplicity, the elements of $K^{(2)} = \hat H^{-1}_{Q_2}$, in the
second block, are
denoted by 
\begin{equation}
K^{(2)}_{\alpha\beta}= \frac{1}{2U_{\alpha\beta}},
\end{equation}
and in the same way, $\alpha, \beta = (1, 2, 3)$, whenever
$\alpha, \beta = (x, y, z)$.

We determine the final form of the effective Hamiltonian by computing
the relevant matrix elements of Eq.~(\ref{H_eff}). To this end, we
consider in detail all the different cases:

The states of the type $\vert \alpha_{\bf i},\alpha_{\bf j}\rangle$,
and the processes
\begin{equation*}
\begin{array}{l}
\hat a^\dagger_{\alpha, \bf i}\hat a_{\alpha, \bf j}\, \hat K\, \hat
a^\dagger_{\alpha, \bf 
j} \hat a_{\alpha, \bf i} \vert\alpha_{\bf i},\alpha_{\bf j}\rangle =
\hat a^\dagger_{\alpha, \bf i}\hat a_{\alpha, \bf j}\, \hat K\sqrt{2}\vert 0,
2\alpha_{\bf j}\rangle\\\\
= \sqrt{2} \hat a^\dagger_{\alpha, \bf i}\hat a_{\alpha, \bf j}\left(
  K^{(1)}_{\alpha\alpha}\vert 0, 2\alpha_{\bf j}\rangle +
  K^{(1)}_{\alpha\beta}\vert 0, 2\beta_{\bf j}\rangle\right)\\\\ =
2K^{(1)}_{\alpha\alpha}\vert \alpha_{\bf i}, \alpha_{\bf j}\rangle,
\end{array}
\end{equation*}
contribute to the effective Hamiltonian with terms of the type
\begin{equation*}
\displaystyle{-\sum_{\langle {\bf i}, {\bf j}\rangle}
\sum_{\alpha,
  \beta, \gamma}\frac{|t^{\alpha}_{{\bf 
    i j}}|^2}{\Lambda}
\left[\left(
  \epsilon^{\alpha\beta\gamma}\right)^2\left( U_{\beta\beta}
  U_{\gamma\gamma} - U^2_{\beta\gamma}\right)\right]\hat n_{\alpha,
\bm i}\hat n_{\alpha, \bm j}}.
\end{equation*}

For the same states, and the processes
\begin{equation*}
\begin{array}{l}
\hat a^\dagger_{\beta, \bf i}\hat a_{\beta, \bf j}\, \hat K\, \hat
a^\dagger_{\alpha, \bf 
j} \hat a_{\alpha, \bf i} \vert\alpha_{\bf i},\alpha_{\bf j}\rangle =
\hat a^\dagger_{\beta, \bf i}\hat a_{\beta, \bf j}\, \hat K\sqrt{2}\vert 0,
2\alpha_{\bf j}\rangle\\\\
=
2K^{(1)}_{\alpha\beta}\vert \beta_{\bf i}, \beta_{\bf j}\rangle,
\end{array}
\end{equation*}
the effective Hamiltonian picks the term
\begin{equation}
\begin{array}{r}
\displaystyle{-\sum_{\langle {\bf i}, {\bf j}\rangle}\sum_{\alpha,
  \beta,\gamma}2\frac{t^{\alpha}_{{\bf 
    i j}}t^{\beta}_{\bf ji}}{\Lambda}\left[\left(\epsilon^{\alpha\beta\gamma}
\right)^ 2 \left(U_{\alpha\beta}U_{\beta\gamma} -
  U_{\alpha\beta}U_{\gamma\gamma}\right)\right]\times} \\[0.5em]
\displaystyle{\hat a^\dagger_{\beta,
\bm i}\hat a_{\alpha, \bm i}\hat a^\dagger_{\beta, \bm j}\hat
a_{\alpha, \bm j}}.
\end{array}
\end{equation}

Next we consider the $\vert\alpha_{\bf i},\beta_{\bf j}\rangle$ states, with
$\alpha\neq\beta$. The processes of the type
\begin{equation}\label{fermions_1}
\begin{array}{l}
\hat a^\dagger_{\alpha, \bf i}\hat a_{\alpha, \bf j}\, \hat K\, \hat a_{\alpha, \bf
j} \hat a_{\alpha, \bf i} \vert\alpha_{\bf i},\beta_{\bf j}\rangle =
\hat a^\dagger_{\alpha, \bf i}\hat a_{\alpha, \bf j}\, \hat K\vert 0,
\alpha_{\bf j}\beta_{\bf j}\rangle\\\\

= \hat a^\dagger_{\alpha, \bf i}\hat a_{\alpha, \bf j}
  K^{(2)}_{\alpha\beta}\vert 0, \alpha_{\bf j}\beta_{\bf j}\rangle =
  K^{(2)}_{\alpha\beta}\vert\alpha_{\bf i},\beta_{\bf j}\rangle, 
\end{array}
\end{equation}
contribute to the effective Hamiltonian with 
\begin{equation*}
\displaystyle{-\sum_{\langle {\bf i}, {\bf
    j}\rangle}\sum_{\alpha,\beta\neq\alpha}|t^{\alpha}_{{\bf 
    i j}}|^2\frac{1}{U_{\alpha\beta}} 
\hat n_{\alpha, \bf i} \hat n_{\beta, \bf j}}.
\end{equation*}

\begin{figure}
\includegraphics[width=0.43\textwidth]{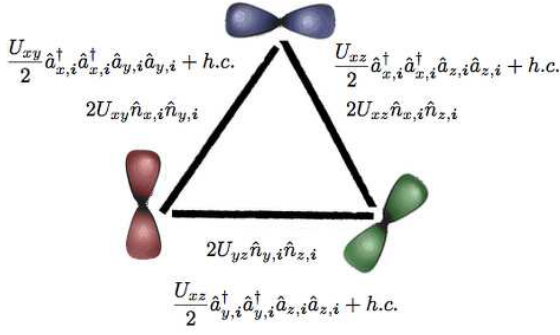}
\caption{(Color online) Dynamical processes relating the different orbital
  states in the many-body bosonic system (as discussed in
  Eq.~(\ref{fermions_mb}), the
  fermionic case contains 
  only the density-density interactions part). Since the different
  orbitals are treated at the same 
  footing, the effective Hamiltonian describing the many-body system in
  the $p$ band with three orbitals requires description in terms of
  the  generators of the $SU(3)$ group. In fact, in the language of
  $p$ orbitals, rather than the triangular scheme displayed above,  
the ladder operators
 of a three-state system with $SU(2)$ symmetry act
  as $p_\alpha \rightleftharpoons p_\beta
 \rightleftharpoons p_\gamma$.}
\label{fig:ladders}
\end{figure}

Finally, for the same states and the processes of the type
\begin{equation}\label{fermions_2}
\begin{array}{l}
\hat a^\dagger_{\beta, \bf i}\hat a_{\beta, \bf j}\, \hat K\, \hat a_{\alpha, \bf
j} \hat a_{\alpha, \bf i} \vert\alpha_{\bf i},\beta_{\bf j}\rangle =
\hat a^\dagger_{\alpha, \bf i}\hat a_{\alpha, \bf j}\, \hat K\vert 0,
\alpha_{\bf j}\beta_{\bf j}\rangle\\\\

= \hat a^\dagger_{\alpha, \bf i}\hat a_{\alpha, \bf j}
  K^{(2)}_{\alpha\beta}\vert 0, \alpha_{\bf j}\beta_{\bf j}\rangle =
  K^{(2)}_{\alpha\beta}\vert\beta_{\bf i}, \alpha_{\bf j}\rangle, 
\end{array}
\end{equation}
the effective Hamiltonian picks the following contribution
\begin{equation*}
\displaystyle{-\sum_{\langle {\bf i}, {\bf j}\rangle}\sum_{\alpha,\beta\neq
  \alpha}t^{\alpha}_{{\bf 
    i j}} t^{\beta}_{\bf ji}\frac{1}{U_{\alpha\beta}}\hat
a^\dagger_{\beta, \bf i}\hat 
a_{\alpha, \bf i}\hat a^\dagger_{\alpha, \bf j}\hat a_{\beta, \bf j}} . 
\end{equation*}
Summing the different contributions, the effective Hamiltonian
describing the 
Mott$_1$ phase of bosons in the $p$ band of the three orbital system
is given by 
\begin{equation}\label{Mott_1-bosons} 
\begin{array}{rcl}
H^b_{M_1} \!\!\!\!\!&=&\!\!\!
\displaystyle{-\!\sum_{\langle {\bf i}, {\bf 
    j}\rangle}\sum_{\alpha,\beta,\gamma}\Bigg[ \frac{|t^{\alpha}_{{\bf 
    ij}}|^2}
{\Lambda}\left( 
  \epsilon^{\alpha\beta\gamma}\right)^2\left( U_{\beta\beta}
  U_{\gamma\gamma} - U^2_{\beta\gamma}\right)
\hat n_{\alpha, \bf i} \hat n_{\alpha,\bf j}} \\[0.5em]
& &\!\!\!\!\!\displaystyle{+ 2\frac{t^{\alpha}_{{\bf 
    i j}}t^\beta_{\bf j i}}{\Lambda}\left(\epsilon^{\alpha\beta\gamma}
\right)^ 2\!\left(U_{\alpha\beta}U_{\beta\gamma} -
  U_{\alpha\beta}U_{\gamma\gamma}\right)
\hat a^\dagger_{\beta, \bf i}\hat a_{\alpha, \bf i}\hat
a^\dagger_{\beta, \bf j}\hat a_{\alpha, \bf j}}\\[0.5em]
& &\!\!\!\!\!\displaystyle{+ \frac{|t^{\alpha}_{{\bf 
    ij}}|^2}{U_{\alpha\beta}} 
\hat n_{\alpha, \bf i} \hat n_{\alpha, \bf j} + \frac{t^{\alpha}_{{\bf 
    i j}} t^{\beta}_{\bf ji}}{U_{\alpha\beta}}\hat
a^\dagger_{\beta, \bf i}\hat 
a_{\alpha,\bf i}\hat a^\dagger_{\alpha,\bf j}\hat a_{\beta, \bf j} \Bigg]}.
\end{array}
\end{equation}
This is a main result of this paper.
We now use the orbital states to define the representation of the
$SU(3)$ group in terms of the 
Gell-Mann matrices $\lambda_i$, $i =
1,..,8$~\cite{greiner1994quantum}. Although individually the 
orbital states have the structure of angular momentum, the generators
of the $SU(2)$ group fail to give a description of the effective
dynamics of the many-body system with three orbitals in the $p$
band. The reason is that dynamical processes in the $p$ band treat any
combination of different orbital states at the same footing. As a
consequence, the ladder operators act as in the Lie algebra of the
$SU(3)$ group (see Fig.~\ref{fig:ladders}).

Using the condition
that $\hat n_{x, \bf i} + \hat n_{y, \bf i} + \hat n_{z, \bf i} = 1$
in the Mott$_1$ phase, the diagonal elements $\lambda_3 = 
\text{diag}(1, -1, 0)$ and $\lambda_8 = 1/\sqrt{3}\,\text{diag(1, 1,
  -2)}$ can be written as  
\begin{equation}\label{gellmann_n}
\begin{array}{rcl}
\hat n_{x,\bf i} &=& \frac{1}{3} + \frac{1}{2}\lambda_{3, \bf i} +
\frac{\sqrt{3}}{6}\lambda_{8, \bf i}\\[0.5em]
\hat n_{y,\bf i} &=&\frac{1}{3} - \frac{1}{2}\lambda_{3, \bf i} +
\frac{\sqrt{3}}{6}\lambda_{8, \bf i}\\[0.5em]
\hat n_{z, \bf i} &=&\frac{1}{3} - \frac{\sqrt{3}}{6}\lambda_{8, \bf i}.
\end{array}
\end{equation}
The $SU(3)$ ladder operators are defined in terms of the non-diagonal
Gell-Mann matrices (see Appendix~\ref{sec:app_su3} for explicit
expressions). In terms 
of the orbital states, 
\begin{equation}\label{ladder_op}
\begin{array}{rcccl}
\frac{\hat T^{\pm}_{\bf i}}{2}  &=& \hat a^\dagger_{x, \bf i}\hat a_{y, \bf
 i} &=& \lambda^{1}_{z, \bf i} \pm i\lambda^{2}_{z, \bf i}   \\[0.5em]
\frac{\hat V^{\pm}_{\bf i}}{2} &=& \hat a^\dagger_{x, \bf i}\hat a_{z, \bf
 i} &=& \lambda^{1}_{y, \bf i} \pm i\lambda^{2}_{y, \bf i} \\[0.5em]
\frac{\hat U^{\pm}_{\bf i}}{2} &=& \hat a^\dagger_{z, \bf i}\hat a_{y, \bf i}
 &=& \lambda^{1}_{x, \bf i} \pm i\lambda^{2}_{x, \bf i}, 
\end{array}
\end{equation}
where we simplified notation by relabelling the Gell-Mann matrices with
the index of the symmetry axis of rotation of the
corresponding $SU(2)$ subalgebra. In the usual setting
$\lambda^1_z = 
\lambda^1$, $\lambda^2_z = \lambda_2$, $\lambda^1_y =
\lambda^4$, $\lambda^2_z = \lambda_5$, $\lambda^1_x =
\lambda^6$ and $\lambda^2_x = \lambda_7$.

\begin{figure}
\includegraphics[width=0.5\textwidth]{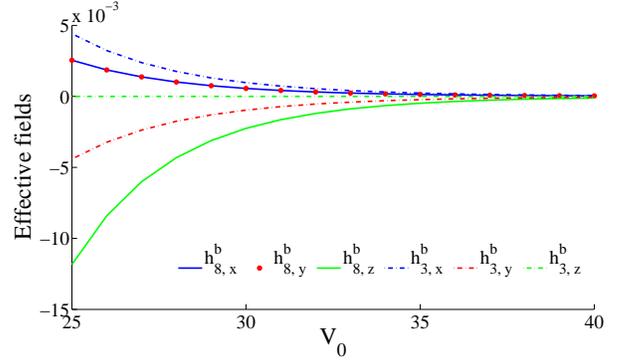}
\caption{(Color online) Effective fields for the bosonic system. The indices labeling
  different directions are used here to illustrate the contribution of
  external field terms in asymmetric lattices (see discussion in
  Sec.~\ref{sec:gs_latt_geo}), i.e., where effective $1D$ and $2D$
  systems are obtained by suppressing the tunneling in 2 or 1
  directions. In particular, due to the symmetries of the dynamics in
  cubic lattices, the term
associated to $h^b_3 = h^b_{3, x} + h^b_{3, y}$ vanishes.}
\label{fig:eff_fields}
\end{figure}

\begin{figure}
\includegraphics[width=0.5\textwidth]{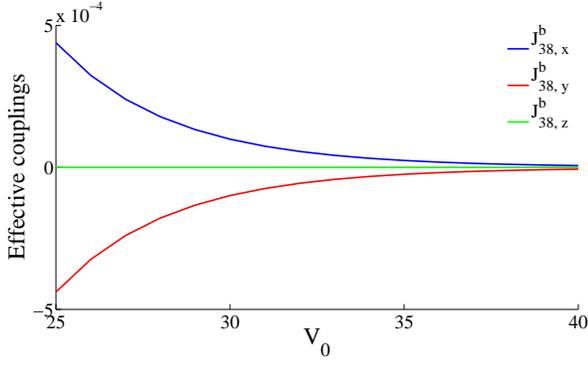}
\caption{(Color online) Effective couplings for the nearest neighbour interactions
  $\left( \lambda_{8, \bm i}\lambda_{3, 
    \bm i} + \lambda_{3, \bm i}\lambda_{8, 
    \bm i} \right)$. In the same way as for the external fields shown
in Fig.~\ref{fig:eff_fields}, the relative sign for the couplings of
dynamics in the $x$ and $y$ directions follow directly from the
symmetries in the $p$ band of isotropic cubic lattices
(see details in the text). In addition, these processes vanish in the
case of isotropic cubic lattices.}
\label{fig:J38}
\end{figure}

\begin{figure}
\includegraphics[width=0.5\textwidth]{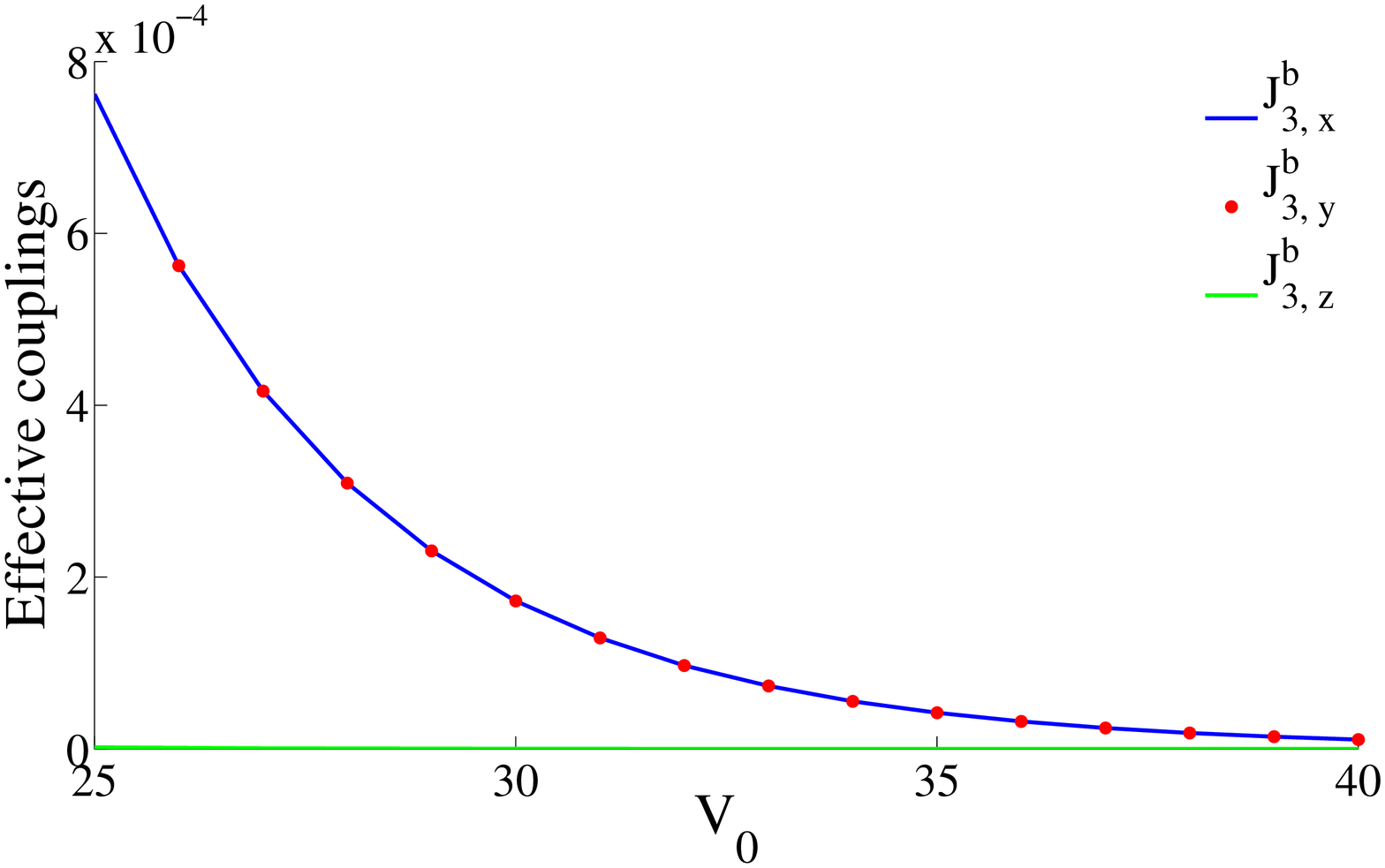}
\caption{(Color online) Effective couplings of the bosonic many-body system for
  $\lambda_{3, \bm i}\lambda_{3, \bm 
    j}$ interactions in the different directions.}
\label{fig:J_3}
\end{figure}

\begin{figure}
\includegraphics[width=0.5\textwidth]{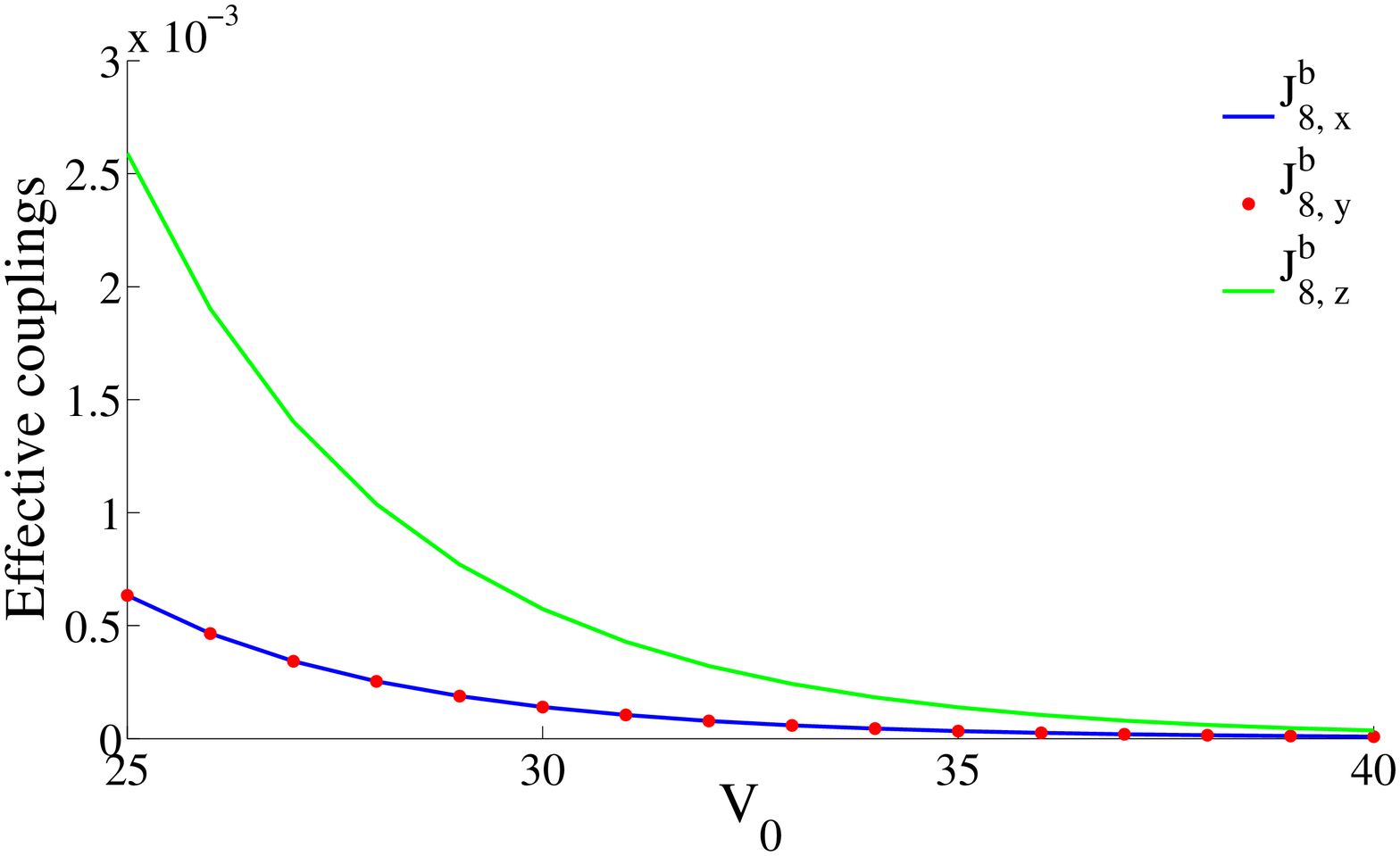}
\caption{(Color online) Effective couplings of the bosonic many-body system for
  $\lambda_{8, \bm i}\lambda_{8, \bm 
    j}$ interactions in the different directions.}
\label{fig:J_8}
\end{figure}

\begin{figure}
\includegraphics[width=0.5\textwidth]{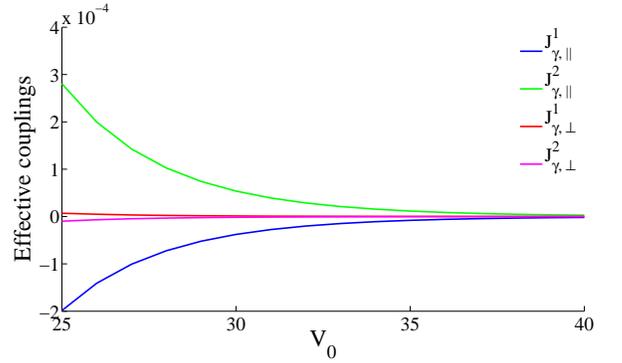}
\caption{(Color online) Effective couplings for nearest neighbour interactions
  obtained from the ladder operators. Notice, in particular, that the
  couplings associated to the $\lambda^1_{\gamma, \bm i}\lambda^1_{\gamma, \bm j}$
  interactions are not the same as the couplings of
  $\lambda^2_{\gamma, \bm i}\lambda^2_{\gamma, \bm j}$. This is a 
  $XY\!Z$-like anisotropy, which is typical for the bosonic
  system in the $p$ band~\cite{pinheiro2013xyz}. In fact, this is a
  direct consequence from the combination of anisotropic tunneling
  with orbital-changing interaction terms.}
\label{fig:anisotr_par}
\end{figure}

This allows the Hamiltonian~(\ref{Mott_1-bosons}) to be written in a more
compact form, 
\begin{equation}\label{eff_H_b}
\begin{array}{rcl}
\hat H^b_{M_1} &=&\displaystyle{-\sum_\sigma\sum_{\langle
  \bf{i},\bf{j}\rangle_\sigma} \Big[ J^b_{3, \sigma}
\lambda_{3,\bf i}\lambda_{3,\bf j} + J^b_{8, \sigma} \lambda_{8,\bf
  i}\lambda_{8,\bf j}} \\[0.5em]
& &\displaystyle{+ J^b_{38, \sigma}\left(\lambda_{3,\bf
   i}\lambda_{8,\bf j} + \lambda_{8,\bf i}\lambda_{
   3,\bf j}\right) }\\[0.5em]
& & \displaystyle{+ \sum_\gamma J^1_{\gamma, \sigma}\lambda^{1}_{\gamma, \bf
    i}\lambda^{1}_{\gamma, \bf j} +J^2_{\gamma,\sigma}\lambda^{2}_{\gamma, \bf
  i}\lambda^{2}_{\gamma,\bf j}\Big]}\\[0.5em]
& &\displaystyle{ -\sum_{\bm i} \Big(h^b_{3}\lambda_{3, \bm i} +
  h^b_{8}\lambda_{8, \bm i} \Big)} . 
\end{array}
\end{equation}
Explicit expressions of the coupling constants are not very
informative at first sight and can be found in
Appendix~\ref{sec:app_cc}. The physics of the system is
summarized here, however, from the study of the couplings as
a function of the lattice depth, as illustrated in
Figs.~\ref{fig:eff_fields} -~\ref{fig:anisotr_par}.  
They are computed from diagonalization of the Mathieu equation for the
potential~(\ref{per_pot}) with $V_\sigma = V_0$, $\sigma = x, y, z$,
which yields the lattice Wannier functions used in the construction of
the $p$ orbitals. In the same way as $V_0$, all the couplings and
external fields are scaled with $E_R$.

First,
due to the tunneling anisotropy, the values of the coupling
constants will depend on the direction of
dynamics. 
This system contains two external fields $h^b_{3}$
and $h^b_{8}$, associated to $\lambda_{3}$ and $\lambda_8$,
that are shown in Fig.~\ref{fig:eff_fields}.
These are, respectively, the isospin and 
hypercharge operators in the study of strong interactions in QCD,
whose eigenvalues are used to 
label the states of the $SU(3)$
multiplet~\cite{greiner1994quantum}. In the context of $p$-band 
physics, the external fields are related to population imbalance in
the different orbital states at each site.
In order to connect this notation with the one used in 
Fig.~\ref{fig:eff_fields}, we notice that $h^b_{8} =\sum_\sigma
h^b_{8, \sigma }$ and $h^b_{3} =\sum_\sigma
h^b_{3, \sigma}$. This choice will become more transparent in the
discussion of realizations of the effective models in $1D$ and $2D$
lattices in 
Sec.~\ref{sec:gs_latt_geo}. For the moment, however, we remark that as
a consequence of the symmetries of the dynamics in the $p$ band of
isotropic cubic lattices, contributions of the 
$h^3_{3}$ external field vanish after summation over $\sigma$.

The leading interaction terms in this Hamiltonian stem from
contributions of density-density interactions in the bosonic
picture, both between atoms in the same and in different orbital
states. Here
again, due
to the symmetries of the dynamics in the $p$ band
$J_{38, x} = -J_{38, y}$ (see Fig.~\ref{fig:J38}), and therefore the
term with coefficient $J^b_{38, 
  \sigma}$ vanishes after summation over $\sigma$ in the case of a
cubic lattice.

Finally, in the same way as for the effective dynamics of the
two-orbital bosonic 
system in the $p$ band in Ref.~\cite{pinheiro2013xyz}, the
nearest-neighbour interactions obtained from the ladder operators
yield $XY\!Z$-like anisotropies in the couplings of $\lambda^1_\gamma$ 
and $\lambda^2_\gamma$. As shown in Fig.~\ref{fig:anisotr_par},
the directions parallel and perpendicular to $\gamma$ will have
different couplings. However, the symmetries of dynamics in the $p$
band of cubic lattices imply 
that $J^1_{\gamma, \parallel}$ and $J^2_{\gamma, \perp}$ are the same
for all values of $\gamma = x, y, z$ (recall that $\gamma$ is the
label of the different $SU(2)$ subalgebras).

\subsection{The effective Hamiltonian for spinless fermions}
\begin{figure}
\includegraphics[width=0.5\textwidth]{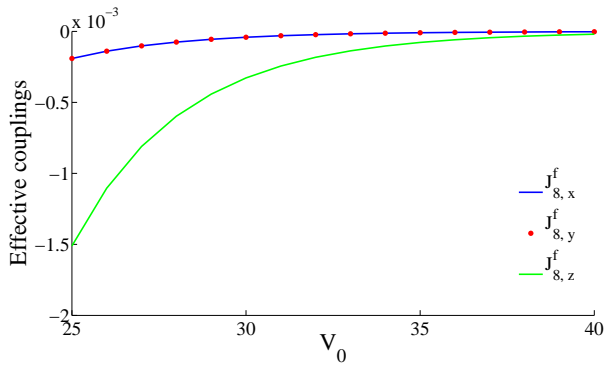}
\caption{(Color online) $J^f_8$ for the fermionic case. Notice here that the
  $\lambda_{8,\bm i}\lambda_{8,\bm 
    j}$ term contributes very little for the dynamics in the $x$ and
  $y$ directions, while it is very significant for the dynamics
  in the $z$ direction.}
\label{fig:f_J8}
\end{figure}

\begin{figure}
\includegraphics[width=0.5\textwidth]{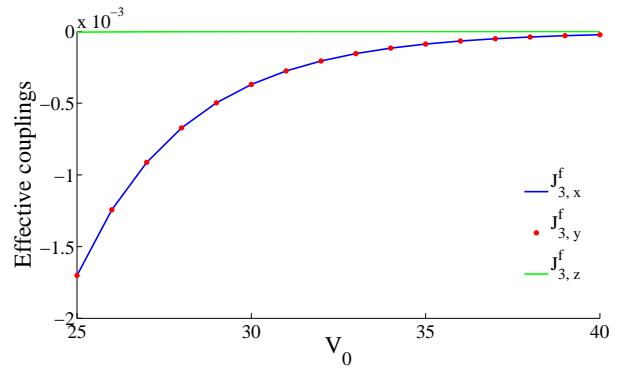}
\caption{(Color online) Coefficients of $\lambda_{3, \bm i}\lambda_{3, \bm j}$
  for interactions in the fermionic case in the different directions
  and as a function of the lattice 
  depth.}
\label{fig:f_J3}
\end{figure}

\begin{figure}
\includegraphics[width=0.5\textwidth]{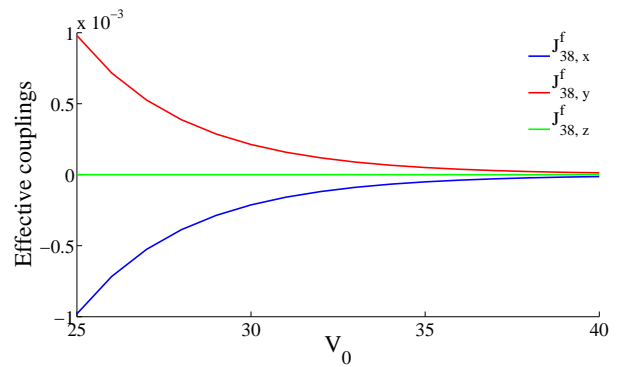}
\caption{(Color online) Coefficients of $\left( \lambda_{3, \bm i}\lambda_{8, \bm j}
    + \lambda_{8, \bm i}\lambda_{\bm, j}\right)$ for the fermionic
  case. In the same way as discussed in the bosonic case, these terms
  yield no contribution to the energy in isotropic cubic lattices. }
\label{fig:f_J38}
\end{figure}

\begin{figure}
\includegraphics[width=0.5\textwidth]{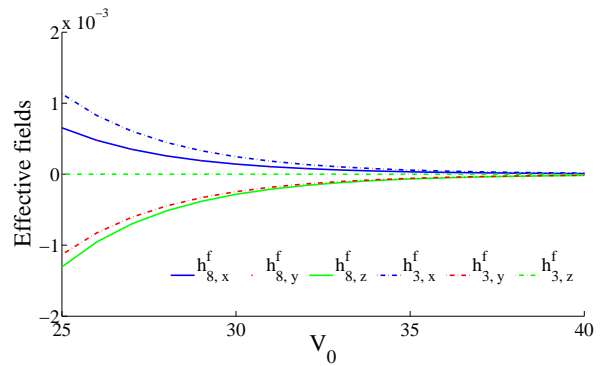}
\caption{(Color online) Effective fields of the effective model with fermionic atoms
  in isotropic cubic lattices. The situation is again similar to what
  is discussed in the bosonic case in Fig.~\ref{fig:eff_fields}. This
  should be the case, since the external fields account for
  single particle contributions and therefore are independent from the
  statistic of the atoms. In the $p$-band system, in particular, the
  external fields encode the degree of imbalance in the occupation of
  the different orbital states.}
\label{fig:f_eff_fields}
\end{figure}

\begin{figure}
\includegraphics[width=0.5\textwidth]{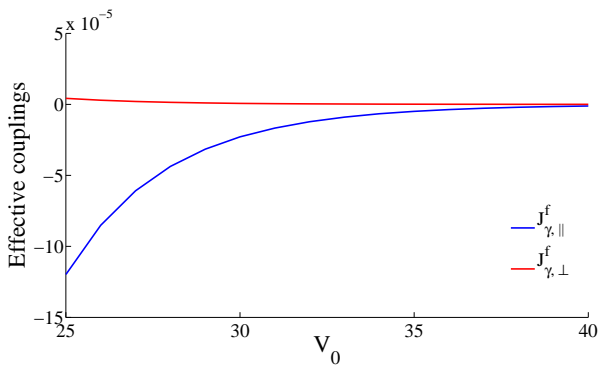}
\caption{(Color online) Coefficients of the interaction stemming from the ladder
  operators in the fermionic case. Notice, in particular, the
  $XXZ$-like type of couplings, which differ from the $XY\!Z$-like
  couplings obtained for the same terms in the bosonic case (see
  Fig.~\ref{fig:anisotr_par}).}
\label{fig:f_ladder_coef}
\end{figure}

The derivation of the effective Hamiltonian describing the Mott$_1$
phase of fermions in the $p$ band of $3D$ lattices is simplified due
to the absence of interactions involving two atoms in the same orbital
state. Here, $\hat H_Q$ is diagonal in the basis of
intermediate states of the perturbative calculation, and the only
processes to take into account are described in
Eqs.~(\ref{fermions_1}) and~(\ref{fermions_2}). Therefore, 
\begin{equation}
\begin{array}{rcl}
\hat H^f_{M_1}&=&\displaystyle{-\sum_{\langle {\bf i}, {\bf
    j}\rangle}\sum_{\alpha,\beta\neq\alpha}\bigg[2|t^{\alpha}_{{\bf 
    i j}}|^2\frac{1}{2U_{\alpha\beta}} 
\hat n_{\alpha, \bf i} \hat n_{\beta, \bf j}} \\[0.5em]
& &\displaystyle{+ 2t^{\alpha}_{{\bf 
    i j}} t^{\beta}_{\bf ji}\frac{1}{2U_{\alpha\beta}}\hat
a^\dagger_{\beta, \bf i}\hat 
a_{\alpha\bf i}\hat a^\dagger_{\alpha\bf j}\hat a_{\beta \bf j}\bigg]},
\end{array}
\end{equation} 
which in terms of the Gell-Mann matrices (see Eqs.~(\ref{gellmann_n})
and~(\ref{ladder_op})) becomes
\begin{equation}\label{eff_H_f}
\begin{array}{rcl}
\hat H^f_{M_1} &=&\displaystyle{-\sum_\sigma\sum_{\langle
  \bf{i},\bf{j}\rangle_\sigma} \Big[ J^f_{3, \sigma}
\lambda_{3,\bf i}\lambda_{3,\bf j} + J^f_{8, \sigma} \lambda_{8,\bf
  i}\lambda_{8,\bf j}} \\[0.5em]
& &\displaystyle{+ J^f_{38, \sigma}\left(\lambda_{3,\bf
   i}\lambda_{ 8,\bf j} + \lambda_{8,\bf i}\lambda_{
   3,\bf j}\right)} \\[0.5em]
& & \displaystyle{+  \sum_{\gamma}J^f_{\gamma, \sigma}\left(\lambda^{1}_{\gamma, \bf
    i}\lambda^{1}_{\gamma, \bf j} + \lambda^{2}_{\gamma, \bf
  i}\lambda^{2}_{\gamma,\bf j}\right)\Big]}\\[0.5em]
& &\displaystyle{ -\sum_{\bm i} \Big(h^f_{3}\lambda_{3, \bm i} + h^f_{8}\lambda_{8,
  \bm i} \Big)}. 
\end{array}
\end{equation}
Following the analysis of the bosonic case, the expressions of the
couplings are given in the Appendix, and the behavior of the
couplings as a function of the lattice depths is shown in
Figs.~\ref{fig:f_J8} -~\ref{fig:f_ladder_coef}. Here too, the leading
terms contain contributions deriving from density-density
interactions in the many-body system, and are 
accordingly mediated by the diagonal Gell-Mann matrices. In the
fermionic case, however, density-density interactions are only
between atoms in different orbital states. In addition, the vanishing
of contributions associated to $h^f_3$ and $J^f_{38,
  \sigma}$ in the cubic lattice has the same explanation already 
discussed for the bosonic case, which is characterized by the same
property.

Further comparison between the effective models derived for the
many-body bosonic and fermionic systems reveal two main
differences. The first one is manifest in the interaction terms
derived from contributions of the ladder operators. Namely, the
fermionic case lacks the $XY\!Z$-like anisotropy in the couplings of
$\lambda^{1}_{\gamma, \bf i}\lambda^{1}_{\gamma, \bf j}$ and $
  \lambda^{2}_{\gamma, \bf i}\lambda^{2}_{\gamma,\bf j}$. In fact, the
  coupling of these terms is much closer in form to the couplings in
  $XXZ$ $SU(2)$ Heisenberg models. The second
  difference is related to 
  the type of ordering preferred for the different systems. While the
  bosonic case favors ferromagnetic alignment of the degrees of
  freedom in the leading terms (with couplings $J^b_{3, \sigma}$ and
  $J^b_{8, \sigma}$), the fermionic case favors antiferromagnetic
  alignment (see Figs.~\ref{fig:J_3},~\ref{fig:f_J3},~\ref{fig:J_8}
  and~\ref{fig:f_J8}).


\subsection{Different lattices geometries and exotic ground states}
\label{sec:gs_latt_geo}
Owing to the complexity of these systems, ground-state 
properties of $SU(3)$ Heisenberg models have been characterized mainly
in systems with isotropic
couplings in $1D$ and $2D$~\cite{papanicolaou1988unusual}. Further
studies of both the 
triangular and the square lattices in $2D$, and the cubic lattice in
$3D$ have been carried out numerically only rather
recently~\cite{toth2010three, Bauer:2012aa}, and complement
the previous analysis based on application of flavor
wave theory to the case with antiferromagnetic
couplings~\cite{papanicolaou1988unusual}. 
These studies confirm the prediction of a highly degenerate manifold of
ground-states with both $2$- and $3$-
sublattice ordering for the square and cubic lattices, whose
degeneracy is lifted by a mechanism of
order-by-disorder~\cite{toth2010three, Bauer:2012aa}.  

This is an example of the many interesting phenomena that can
be experimentally explored with the systems discussed here. In fact, since 
the Hamiltonian Eq.~(\ref{eff_H_f}) corresponds to that of 
an $XXZ$-like $SU(3)$ Heisenberg model, the $XXZ$-type of anisotropy
will most likely lead to a rich phase diagram for the corresponding
ground state, as is the case in  
$SU(2)$ Heisenberg models (see~\cite{mikeska2004one}, for example). 
The flavor-wave analysis of this system in the $3D$ lattice is however
left for the future.

By increasing the depth of the lattice sites in one of the directions,
say $z$, so as to suppress the tunneling but still keeping the
degeneracy between the different orbital states, it is possible to
study the physics of $2D$ lattices~\cite{li2012time}. As reported
in~\cite{toth2010three, Bauer:2012aa}, the physics here should include
the formation of $2$- and $3$-sublattice ordering, with preference for
the $2$-sublattice ordering at energies below the energy scale of
exchange processes. By increasing the
depth of the tunneling in two directions while still keeping the
degeneracy between the three orbital states, different
types of chains can then be obtained. For example, if the tunneling is
restricted to the 
$z$-direction, the effective Hamiltonian will not contain contributions
from $h^f_{3}$, and the contributions stemming from the term with the
$J^f_{38, \sigma}$ coupling will cancel due to symmetry. By allowing the
dynamics in only $x$ or only $y$ directions, the effective Hamiltonian
will pick contributions of all terms and it is therefore possible to
study antiferromagnetic $XXZ$-like $SU(3)$ models in external fields.

Ferromagnetic $SU(3)$ Heisenberg models can also be engineered in $1D$
and $2D$ by considering the bosonic system, and suppressing the
dynamics in 2 or 1 directions, respectively. The main difference, 
however, is  the $XY\!Z$ type of anisotropy in the couplings derived
from the ladder operators. The same as in the fermionic case, and in
analogy to the situation in $SU(2)$ Heisenberg models, this 
coupling anisotropy is expected to give rise to very rich physics. To
leading order, an effective $2D$ 
system obtained from suppressing the dynamics in the $x$ or $y$
direction, for example, will contain contributions of both the term of
the free field $h^b_{3}$ and of the interaction term with coupling
$J^b_{38}$. This scenario is closer to that recently
reported in Ref.~\cite{grass2014spiral}, which predicts the existence
of spiral spin textures in the ground-states of ferromagnetic $SU(3)$
Heisenberg models. In the $p$ band system, this
would be cast as a rotation (or change in the relative angle of the
onsite orbital orientations) of one of the orbital states at each
lattice site. We expect that similar physics will appear in the $1D$
case.

\section{Experimental probes}\label{sec:exp_prob}
The physics discussed here takes place deep in the Mott insulator phase,
where single sites of the optical lattice can be accurately
approximated by a harmonic potential with frequency 
$\omega_\alpha = \sqrt{2V_\alpha 
  k_\alpha^2/m}$ (recall that $k_{\alpha}$ are the wave vectors of the
optical lattice laser in the direction
$\alpha$)~\cite{haroche2013exploring}. Different vibrational 
levels in this potential, which in the context of the optical lattices
correspond to the different bands, can then be coupled by performing
stimulated Raman transitions in a two-level atom~\cite{muller2007state}. This 
technique has been employed in Ref.~\cite{muller2007state} for
promoting atoms from 
the $s$ to $p$ bands, 
in the Mott phase, of $1D$, $2D$ and $3D$ lattices. However, since the
main object of that study was concerned with the properties of
coherence, i.e., the 
superfluid phase, further manipulation of orbital degrees of freedom
in the Mott phase have not been discussed.  We therefore extend the
methods introduced in~\cite{pinheiro2013xyz} for control and
manipulation of $SU(2)$ Heisenberg models based on orbital degrees of
freedom, for the implementation of $SU(3)$ Heisenberg models.

Consider a Raman coupling between the $\vert 1\rangle = \vert F =
1\rangle $ and $\vert 2 
\rangle = \vert F = 2\rangle$
atomic electronic
states of $^{87}$Rb. These are two-photon
processes where the two levels are coupled with
an intermediate virtual state, far detuned from all the other
states of the system~\cite{muller2007state}. Because of this intermediate coupling,
implementation of Raman transitions require the use of two
different lasers, whose corresponding wave vectors are denoted here
by $\vec k_{L_1}$ and $\vec k_{L_2}$.
The matrix element
characterizing this transition is given by 
\begin{equation}
\frac{\Omega_1\Omega_2^{*}}{\delta}\langle 2\vert e^{i (\vec
  k_{L_1} - \vec k_{L_2}).\vec x }\vert 1\rangle, 
\end{equation}
where $\Omega_i$ are the Rabi frequencies between the $\vert i\rangle$
states, $i = 1, 2$ with another far detuned auxiliary state of this
system, say 
$\vert aux\rangle$, and 
$\delta$ is the detuning between $\vert aux\rangle$ and the
virtual intermediate state.   

\begin{figure}
\includegraphics[width=0.285\textwidth]{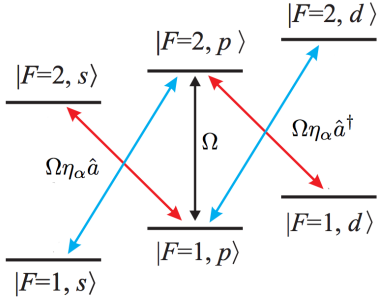}
\caption{(Color online) Schematic couplings between the different orbital
  states. While the carrier transition does not change the vibrational
state of the atom, red and blue sideband transitions can be used to
lower and raise the vibrational states of the atom, which therefore couples
different orbital states.}
\label{fig:ion_su3}
\end{figure}

After adiabatic elimination of the auxiliary state, the interaction
between the atom in the harmonic potential with the   
lasers driving the Raman coupling is given
by~\cite{haroche2013exploring} 
\begin{equation}
H = \sum_{\alpha} \omega_{\alpha} \hat a^\dagger a -
\sum_{\alpha}\left[\frac{\Delta_\alpha}{2}\sigma_z + 
\frac{1}{2}\Omega\left( \sigma_+e^{i\eta_{\alpha}(\hat a + \hat a^\dagger)} +
  h. c.\right)\right],
\end{equation}
where first term accounts for the center of mass motion of the atom
in the harmonic potential, and the second and
third terms describe the driven two-level system in the rotating-wave
approximation~\cite{haroche2013exploring} .  
In this notation $\Omega = \Omega_1\Omega_2^{*}/\delta$ is the
effective Rabi frequency, $\sigma_+ = (\sigma_-)^\dagger =
\vert 2\rangle\langle 1\vert$, $\sigma_z = \vert 2\rangle \langle
2\vert - \vert 1\rangle \langle 1\vert$, 
$\Delta_\alpha = \omega_\alpha - \omega_{12}$ are
the detunings of the lasers with respect to
the atomic transition, of frequency $\omega_{12}$, and $\eta_{\alpha}
= \Delta k_{L,\alpha}\sqrt{\hbar/2m\omega_\alpha}$ is the Lamb-Dicke
parameter, with $\Delta k_{L, \alpha} = k_{L_1, \alpha} - k_{L_2, \alpha}$.

In the Lamb-Dicke regime, when $\eta_\alpha \ll 1$, the expansion of
the exponential can be truncated to $e^{i\eta_\alpha(\hat a + \hat
  a^\dagger)} \approx 1 + \eta_\alpha(\hat a + \hat
a^\dagger)$~\cite{haroche2013exploring}, and the corresponding
Hamiltonian describes    
a two-level system coupled to the
phonon excitations of the harmonic oscillator with bare Hamiltonian
given by $H_0 = \hat a^\dagger a
- \frac{1}{2}\Delta_\alpha\sigma_z$. The eigenstates of this system
can be denoted by $\vert 1, n\rangle$ and $\vert 2,
n\rangle$, where $n$ labels the vibrational level. By carefully
choosing the driver
frequency, three possible transitions 
can be implemented~\cite{haroche2013exploring}:
The carrier transition, when $\Delta_\alpha = 0$,
\begin{equation}
H_{car} = \frac{\hbar}{2}\Omega\left [\sigma_+ + h.c.\right],
\end{equation}
which has no effect in the vibrational state. By choosing $\Delta_\alpha =
-\omega_\alpha$, the red sideband transitions 
\begin{equation}
H_{rsb} = \frac{\hbar}{2}\Omega\eta_\alpha\left [\hat a
  (\sigma_-)^\dagger + \hat a ^\dagger \sigma_-\right] 
\end{equation}
decrease the vibrational state $n$ by one quanta, when the atom swaps
from $\vert 1\rangle$ to $\vert 2 \rangle$.
Finally, when  
$\Delta_\alpha = \omega_\alpha$ one implement blue sideband
transitions,
\begin{equation}
H_{bsb} = \frac{\hbar}{2}\Omega \eta_\alpha \left[\hat
  a^\dagger(\sigma_-)^\dagger + \hat a \sigma_- \right],
\end{equation}
 that increase the vibrational level $n$ by one for the same atomic
 transition.   
 These transitions are
 schematically shown in 
Fig.~\ref{fig:ion_su3}.  In addition to selective transitions, it is also
possible to 
selectively address the different orbital
states~\cite{bakr2009quantum}. $p_x$ orbitals, for 
example, can be addressed by choosing
driver lasers with no 
component in the $y$ and $z$ directions, i.e, $\vec k_{L_1} - \vec
k_{L_2} = k_{L_x}$. Analogous relations hold for manipulations of only
$p_y$ and/or $p_z$ orbitals.

Finally, we show that these (trapped ion based) techniques allow for
complete control of these systems by discussing the schemes for
implementation of arbitrary rotations. They are performed here by the
generators of the $SU(3)$ group, and for a given effective angle $\phi$,
$\hat R_\beta(\phi) = 
e^{-i\lambda_\beta\phi/2}$, with $\beta = 1,...,8$. The simplest case, of
rotations with 
$\lambda_8$, can be achieved via Stark shift of the $p_z$ orbital
without any disturbance of $p_x$ and $p_y$ orbitals. $\hat R_3(\phi)$
rotations are also implemented via Stark shift, but now with a
dispersive coupling between both $p_x$ and $p_y$ orbitals which
already renders the shift with correct (opposite) sign. $\hat R_1(\phi)$,
$\hat R_4(\phi)$ and $\hat R_6(\phi)$ rotations are implemented by
driving red sideband transitions off-resonantly, for two
orbitals. The first case will involve the $p_x$ and $p_y$ orbitals,
while in the second and third, $p_x$ and $p_z$, and $p_y$ and $p_z$
orbitals, respectively. The other three rotations, around $\lambda_2$,
$\lambda_5$ and $\lambda_7$ can be achieved by noticing that 
$\hat R_2(\phi) = \hat R_3(\pi/2)\hat R_1(\phi)\hat R_3(-\pi/2)$,
$\hat R_5(\phi) = \hat R_3(\pi)\hat R_8(-\sqrt{3}\pi)R_4(\phi)
\hat R_3(-\pi)\hat R_8(\sqrt{3}\pi)$, and 
$\hat R_7(\phi) = \hat
R_3(\pi/2)\hat R_8(\sqrt{3}\pi/2)R_6(\phi) 
\hat R_3(-\pi/2)\hat R_8(- \sqrt{3}\pi/2)$. Furthermore, the
states of pseudospins can be resolved with single-site 
fluorescence after measurement of $\lambda_3$ for  $p_x$ and
$p_y$, and of $\lambda_8$ for the $p_z$ orbital. Likewise, 
$\langle\lambda_\alpha\lambda_\beta\rangle$ correlation functions can
be obtained by using the same techniques, but now combined with
coincident measurement of the fluorescent photons.  

Two final remarks regarding experimental feasibility are now in
order. First, that as reported in Ref.~\cite{muller2007state}, after
the process of loading, the atoms in the $p$ band remain in a metastable
state. The leading decaying channel here 
stem from atom-atom collisions~\cite{isacsson2005multiflavor}, and
therefore the conditions of the 
Mott$_1$ phase should corroborate to extending the lifetimes of these
systems.  
Typical lifetimes are estimated to be $\sim
5\,m$s~\cite{pinheiro2013xyz}. This is of the order or
dozens tunneling times and should therefore allow for experimental
probing of the properties discussed here. Second, that although the
temperatures required for observation pseudospin correlations derived
from exchange interaction are of the order of  $k_B T \lesssim
t^\alpha t^\beta/U_{\alpha\alpha} \sim 10^{-9}$K, the increased tunneling rates of
the $p$ band may offer slight advantage as compared to the realizations
proposed with atoms in the ground band. Nevertheless, these
temperatures are still in the frontier achieved by current experiments. 

\section{Imperfections due to the loading to the $p$
  band}\label{sec:imperfec} 
Another challenge with implementation of the effective
models is related to experimental imperfections in the process of
loading atoms to the $p$ band. The techniques used in
Ref.~\cite{muller2007state}, based on 
stimulated Raman transitions reported $80\%$ fidelity in promoting
the atoms from the $s$ band in the Mott phase. Therefore it is
important to understand how the presence of residual $s$-orbital atoms
affect the physics of the system.

Let us start by considering 
\begin{equation}
U_{sp_{\alpha}} = U_0 \int d\vec r |\psi_{s\bm j}(\vec
r)|^2|\psi_{\alpha \bm j}(\vec r)|^2,
\end{equation}
which characterize the strength of repulsive interactions between an
$s$- and a $p_\alpha$-orbital atom at the site $\bm j$. Since the
$p_\alpha$-orbital wave functions are spatially broader than the
$s$-orbital wave functions, $U_{sp_\alpha} > U_{\alpha\beta}$, which
makes the repulsive interaction between $s$-orbital and $p$-orbital
atoms larger than the repulsive interaction when both atoms are in the
$p$ band. Now two additional processes should be considered in the
effective model. The first one, which includes tunneling of
$s$-orbital atoms, can be safely neglected due to the reduced rate of
tunneling in the $s$ band and the larger value of the coupling
constant for 
repulsive interaction $U_{sp}$. The second process, which involves
tunneling of $p_\alpha$-orbital atoms, will contribute to the
Hamiltonian with the following term 
\begin{equation}
-\frac{|t^\alpha_{\bf ij}|^2}{U_{sp}}\hat a_{\alpha, \bm i}^\dagger \hat
a_{s,\bm j}^\dagger \hat a_{\alpha,\bm i}\hat a_{\alpha, \bm j} =
-\frac{|t^\alpha_{\bf ij}|^2}{U_{sp}}\hat n_{\alpha, \bm i},
\end{equation} 
where $\hat a _{s, \bm j}$ $(\hat a_{s, \bm j}^\dagger)$ annihilate
(create) an $s$-orbital atom at the site $\bm j$ and where we used that
$\hat n_{s, \bm j} = 1$. The presence of residual $s$-orbital atoms is
therefore associated with local fluctuations of the external fields,
which in this model are represented by the 
$\lambda_{3, \bm j}$ and $\lambda_{8, \bm j}$ 
Gell-Mann matrices. This means that the effective Hamiltonian contains
the additional term 
\begin{equation}\label{H_dis}
\begin{array}{rcl}
H_{dis} &=&\frac{\sqrt{3}}{6U_{sp}}\lambda_{8\bf, i}\left[|t^x_{\bf
      ij}|^2 + |t^y_{\bf ij}|^2 -2|t^z_{\bf ij}|^2\right] \\[0.5em]
& &+ \frac{\lambda_{3,\bm
   i}}{2U_{sp}}\left[|t^x_{\bf 
      ij}|^2 - |t^y_{\bf ij}|^2\right].
\end{array}
\end{equation}
Notice, here, that in the same way that the coefficients of
the external fields derived in Eqs.~(\ref{eff_H_b})
and~(\ref{eff_H_f}) were direction dependent, the tunneling anisotropy
implies a direction-dependent local shift of the external fields due
to $s$-atom impurities. Isotropic cubic latices, however, have
vanishing contributions of the $\lambda_{3, \bm i}$ terms. 

Since the loading of atoms to the $p$ band is implemented globally, a
coherent loading will prepare translationally invariant
states with a fraction of the population in the $s$ band. However,
whenever the loading is not perfectly coherent,
we may envision situations where decoherence process lock the $s$-band
atoms at fixed sites. In such cases, the collapse of the state
describing these residual atoms, induced by decoherence, will break the
translational symmetry and the overall effect of the $s$-band
atoms will be that of a static disorder in the fields as in Eq.~(\ref{H_dis}).
In light of the Imry-Ma argument~\cite{imry-ma1975random}, which
establishes 
a criteria for the  
stability of ordered phases in the presence of disorder, we expect the
presence of $s$-atom impurities to have a larger effect in the fermionic
system and for effective dynamics in $1D$ and $2D$. The presence of
discrete symmetries in 
the bosonic case should attenuate the effects of impurities even in
lower dimensions, and in this case we expect the physics discussed
here to be robust even in the presence of $s$-orbital atoms.

\section{Conclusions}\label{sec:conclusions}
In this paper we presented an alternative controllable system for
implementation of different Hamiltonians describing nearest-neighbour
interactions between degrees of freedom that are the generators of the
$SU(3)$ group. As explained in the text, the properties of the
couplings of the effective models 
 are determined by the statistics of the atoms. Namely, while the
 effective models 
 associated to the many-body fermionic system contains $XXZ$-type of
 couplings, the effective model associated with the bosonic systems
 features $XY\!Z$-like anisotropies. By suppressing the dynamics in
 one or two directions while keeping the degeneracy of the orbital
 states, it is possible, in addition, to engineer different types of
 $2D$ and $1D$ models.  

The systems discussed here allow for quantum simulation of various
types of $SU(3)$ Heisenberg models. Properties of the ground state of
the antiferromagnetic case with isotropic couplings have been
discussed in the literature, and give rise to very rich physics. This
include, for example, the mechanism of order-by-disorder for lifting
the degeneracy of a highly degenerate manifold of ground
states~\cite{Bauer:2012aa, toth2010three, papanicolaou1988unusual}. In 
analogy to the situation in $SU(2)$ 
Heisenberg models, we expect both the antiferromagnetic 
$XXZ$-like case and the ferromagnetic case with $XY\!Z$-type of
couplings to display a 
rich phase diagram for the ground state. However, characterization of
these properties via flavor-wave analysis in the $2D$ and $3D$ cases
is left for the future. 

As a final remark we notice that the perturbative method discussed in 
Sec.~\ref{sec:eff_ham} combined with the experimental probing of
Sec.~\ref{sec:exp_prob} can be used as a general tool for deriving 
 quantum simulators of exotic models of magnetism. In this framework,
 where the pseudospin degree of freedom is encoded in the 
 orbital states
 of excited bands of optical lattices, properties of the couplings
 mediating interactions in 
 the corresponding effective
 model will intrinsically depend on the dynamical processes featured in the
 many-body system. In this sense, different lattice geometries as
 e.g. the triangular lattice could lead to interesting effects,
 specially via the presence of Dzyaloshinskii-Moriya
 exchange. Furthermore, experimentally controllable $SU(3)$ systems
 might be of interest for realization of topological states of 
 matter~\cite{barnett20123}. 

\section{Acknowledgments}
I thank Jonas Larson, Jani-Petri Martikainen, Stephen Powell, Tobias Grass,
Thomas Quella and Daniele Marmiroli for helpful discussions. Jonas Larson and
Jani-Petri Martikainen are once more acknowledged for comments on the
manuscript. The research has been supported by the Swedish research council
(VR). This paper is dedicated to A. F. R. de
Toledo Piza, on the occasion of his 76$^\text{th}$ birthday.

\onecolumngrid
\appendix

\section{SU(3) symmetry}\label{sec:app_su3}
The SU(3) group has 8 generators, denoted here by $\lambda_i$, $i =
1..8$. Explicit expressions (taken from
Ref.~\cite{greiner1994quantum}) are given below, together with the 
notation used in the text:
\begin{equation*}
\lambda^1_z = \lambda_1 = \left(\begin{array}{ccc}
0 & 1 & 0\\
1 & 0 & 0\\
0 & 0 & 0
\end{array}\right),\,\,\,\,\lambda^2_z =\lambda_2= \left(\begin{array}{ccc}
0 & -i & 0\\
i & 0 & 0\\
0 & 0 & 0
\end{array}\right),\,\,\,\,\lambda^1_y =\lambda_4 = \left(\begin{array}{ccc}
0 & 0 & 1\\
0 & 0 & 0\\
1 & 0 & 0
\end{array}\right),\,\,\,\,\lambda^2_y = \lambda_5 =\left(\begin{array}{ccc}
0 & 0 & -i\\
0 & 0 & 0\\
i & 0 & 0
\end{array}\right),
\end{equation*}
\begin{equation}
\lambda^1_x = \lambda_6 = \left(\begin{array}{ccc}
0 & 0 & 0\\
0 & 0 & 1\\
0 & 1 & 0
\end{array}\right),\,\,\,\,\lambda^2_x =\lambda_7= \left(\begin{array}{ccc}
0 & 0 & 0\\
0 & 0 & -i\\
0 & i & 0
\end{array}\right),\,\,\,\,\lambda_3= \left(\begin{array}{ccc}
1 & 0 & 0\\
0 & -1 & 0\\
0 & 0 & 0
\end{array}\right),\,\,\,\,\lambda_8 =\frac{1}{\sqrt{3}}\left(\begin{array}{ccc}
1 & 0 & 0\\
0 & 1 & 0\\
0 & 0 & -2\end{array}\right).
\end{equation}
The Lie algebra of $SU(3)$ is given by $[\lambda_i, \lambda_j] =
2if_{ijk}\lambda_k$, where $i, j = 1,..,8$ $f_{ijk} = -f_{jik} = -f_{ikj} ..$ are
totally antisymmetric structure constants. The values for the
different combinations of indices follow
$f_{123} = 1$, $f_{147} = -f_{156} = f_{246} = f_{257} = f_{345} =
-f_{367} = \frac{1}{2}$ and $f_{458} = f_{678} = \frac{\sqrt{3}}{2}$.\\

The $SU(3)$ group has two Casimir operators, 
\[
\begin{array}{rcl}
C_1(\lambda_i) &=& \frac{1}{4}\sum_i \lambda_i^2\,\,\,\,\,\text{and}\\\\
C_2(\lambda_i) &=& \frac{1}{8}\sum_{ijk} d_{ijk}\lambda_i\lambda_j\lambda_k,
\end{array}
\]
where 
$d_{118} = d_{228} = d_{338} = -d_{888} = \frac{1}{\sqrt{3}}$, 
$d_{146} = d_{157} = d_{344} = d_{355} = -d_{366} = -d_{377} =
\frac{1}{2}$, and 
$d_{448} = d_{558} =
d_{668} = d_{778} = -\frac{1}{2\sqrt{3}}$.

\section{Coupling constants}\label{sec:app_cc}
The expressions of the various coupling constants used in the text are
given below for both the bosonic and fermionic 
many-body systems. 

\subsection{Bosonic case}
In the notation below we use $\sigma$ to denote $\langle \bm i, \bm
j\rangle_{\sigma}$. This defines the values of the tunneling
amplitudes, which are different for various orbital states in the
different directions. 
\begin{equation}
J^b_{8, \sigma} = \frac{K^{(1)}_{xx}}{3}|t^x_{\bf ij}|^2 +
\frac{K^{(1)}_{yy}|}{3} t^y_{\bf ij}|^2 + \frac{4}{3} K^{(1)}_{zz}|t^z_{\bf
  ij}|^2 + \frac{K^{(2)}_{xy}}{6}\left( |t^x_{\bf ij}|^2 +
  |t^y_{\bf ij}|^2\right) -  \frac{2}{9}K^{(2)}_{xz}\left( |t^x_{\bf ij}|^2 +
  |t^z_{\bf ij}|^2\right) - \frac{2}{9}K^{(2)}_{yz}\left( |t^y_{\bf ij}|^2 +
  |t^z_{\bf ij}|^2\right)  
\end{equation}

\begin{equation}
J^b_{3, \sigma} = K^{(1)}_{xx}|t^x_{\bf ij}|^2 + K^{(1)}_{yy}|t^y_{\bf ij}|^2 -
\frac{K^{(2)}_{xy}}{2}\left( |t^x_{\bf ij}|^2 + 
  |t^y_{\bf ij}|^2\right) 
\end{equation}

\begin{equation}
J^b_{38, \sigma} = \frac{\sqrt{3}}{3}K^{(1)}_{xx}|t^x_{\bf ij}|^2 -
\frac{\sqrt{3}}{3}K^{(1)}_{yy}|t^y_{\bf ij}|^2 
-\sqrt{3}\frac{K^{(2)}_{xz}}{6}\left( |t^x_{\bf ij}|^2 +
  |t^z_{\bf ij}|^2\right) + \sqrt{3}\frac{K^{(2)}_{yz}}{6}\left( |t^y_{\bf ij}|^2 +
  |t^z_{\bf ij}|^2\right)
\end{equation}

\begin{equation}
J^1_{\gamma, \sigma} = K^{(1)}_{\alpha\beta}t^\alpha_{\bf
  ij}t^\beta_{\bf ij} + 2K^{(2)}_{\alpha\beta}t^\alpha_{\bf
  ij}t^\beta_{\bf ij}
\end{equation}

\begin{equation}
J^2_{\gamma, \sigma} = K^{(1)}_{\alpha\beta}t^\alpha_{\bf
  ij}t^\beta_{\bf ij} - 2K^{(2)}_{\alpha\beta}t^\alpha_{\bf
  ij}t^\beta_{\bf ij}
\end{equation}

\begin{equation}
h^b_{8, \sigma} = 4\frac{\sqrt{3}}{9}K^{(1)}_{xx}|t^x_{\bf ij}|^2 +
4\frac{\sqrt{3}}{9}K^{(1)}_{yy}|t^y_{\bf ij}|^2 -
8\frac{\sqrt{3}}{9}K^{(1)}_{zz}|t^z_{\bf ij}|^2 -
\frac{\sqrt{3}}{9}K^{(2)}_{xy}\left( |t^x_{\bf ij}|^2 + 
  |t^y_{\bf ij}|^2\right) - \frac{\sqrt{3}}{9}K^{(2)}_{yz}\left( |t^y_{\bf ij}|^2 + 
  |t^z_{\bf ij}|^2\right)
\end{equation}

\begin{equation}
h^b_{3, \sigma} = \frac{4}{3}K^{(1)}_{xx}|t^x_{\bf ij}|^2 -\frac{4}{3}K^{(1)}_{yy}|t^y_{\bf ij}|^2 + 
\frac{K^{(2)}_{xz}}{3}\left(|t^x_{\bf ij}|^2 +
 |t^z_{\bf ij}|^2\right) 
- \frac{K^{(2)}_{yz}}{3}\left( |t^y_{\bf ij}|^2 +
  |t^z_{\bf ij}|^2\right) 
\end{equation}

\subsection{Fermionic case}
In the same way as for the bosonic case discussed above, $\sigma$ is
used below to define the value of the tunneling amplitudes. 
\begin{equation}
J^f_{8, \sigma} = \frac{K^{(2)}_{xy}}{6}\left( |t^x_{\bf ij}|^2 +
  |t^y_{\bf ij}|^2\right) - 2\frac{K^{(2)}_{xz}}{9}\left( |t^x_{\bf ij}|^2 +
  |t^z_{\bf ij}|^2\right) - 2\frac{K^{(2)}_{yz}}{9}\left( |t^y_{\bf ij}|^2 +
  |t^z_{\bf ij}|^2\right)  
\end{equation}

\begin{equation}
J^f_{3, \sigma} = -\frac{K^{(2)}_{xy}}{2}\left( |t^x_{\bf ij}|^2 +
  |t^y_{\bf ij}|^2\right) 
\end{equation}

\begin{equation}
J^f_{38, \sigma} = -\sqrt{3}\frac{K^{(2)}_{xz}}{6}\left( |t^x_{\bf ij}|^2 +
  |t^z_{\bf ij}|^2\right) + \sqrt{3}\frac{K^{(2)}_{yz}}{6}\left( |t^y_{\bf ij}|^2 +
  |t^z_{\bf ij}|^2\right)
\end{equation}

\begin{equation}
J^f_{\gamma, \sigma} = t^\alpha_{\bf ij}t^\beta_{\bf ji} K^{(2)}_{\alpha\beta}
\end{equation}

\begin{equation}
h^f_{8, \sigma} = 2\frac{\sqrt{3}}{9}K^{(2)}_{xy}\left( |t^x_{\bf ij}|^2 +
  |t^y_{\bf ij}|^2\right) - \frac{\sqrt{3}}{9}K^{(2)}_{xz}\left( |t^x_{\bf ij}|^2 +
  |t^z_{\bf ij}|^2\right) - \frac{\sqrt{3}}{9}K^{(2)}_{yz}\left( |t^y_{\bf ij}|^2 +
  |t^z_{\bf ij}|^2\right)
\end{equation}

\begin{equation}
h^f_{3, \sigma} = \frac{K^{(2)}_{xz}}{3}\left( |t^x_{\bf ij}|^2 +
  |t^z_{\bf ij}|^2\right) - \frac{K^{(2)}_{yz}}{3}\left( |t^y_{\bf ij}|^2 +
  |t^z_{\bf ij}|^2\right) 
\end{equation}

\twocolumngrid
\bibliographystyle{ieeetr}
\bibliography{references_pra}


\end{document}